\documentclass[amsmath,amssymb,superscriptaddress,nobalancelastpage,prl,twocolumn,showpacs]{revtex4}
\usepackage{graphicx}
\usepackage{color}
\begin{document}

\title{Field-induced soft-mode quantum phase transition in La$_{1.855}$Sr$_{0.145}$CuO$_{4}$}

\author{J.\ Chang}
\affiliation{Laboratory for Neutron Scattering, ETH Zurich and PSI Villigen, CH-5232 Villigen PSI, Switzerland}

\author{N. B.\ Christensen}
\affiliation{Laboratory for Neutron Scattering, ETH Zurich and PSI Villigen, CH-5232 Villigen PSI, Switzerland}
\affiliation{Materials Research Department, Ris\o\ National Laboratory for Sustainable Energy, DK-4000 Roskilde, Denmark}
\affiliation{Nano-Science Center, Niels Bohr Institute, University of Copenhagen, DK-2100 Copenhagen, Denmark}

\author{Ch.\  Niedermayer}
\affiliation{Laboratory for Neutron Scattering, ETH Zurich and PSI Villigen, CH-5232 Villigen PSI, Switzerland}

\author{K.\ Lefmann}
\affiliation{Materials Research Department, Ris\o\ National Laboratory for Sustainable Energy, DK-4000 Roskilde, Denmark}
\affiliation{Nano-Science Center, Niels Bohr Institute, University of Copenhagen, DK-2100 Copenhagen, Denmark}

\author{H. M.\ R\o nnow}
\affiliation{Laboratory for Neutron Scattering, ETH Zurich and PSI Villigen, CH-5232 Villigen PSI, Switzerland}
\affiliation{Laboratory for Quantum Magnetism, \'Ecole Polytechnique F\'ed\'erale de Lausanne (EPFL), CH-1015 Lausanne, Switzerland}

\author{D. F.\ McMorrow}
\affiliation{London Centre for Nanotechnology and Department of Physics and Astronomy,  University College London, London, UK}

\author{A.\ Schneidewind}
\affiliation{Institut f\"ur Festk\"orperphysik, Technische Universit\"at Dresden, D-01062
Dresden, Germany}
\affiliation{Forschungsneutronenquelle Heinz-Maier-Leibnitz (FRM-II), TU  M\"{u}nchen, D-85747 Garching, Germany}

\author{P.\ Link}
\affiliation{Forschungsneutronenquelle Heinz-Maier-Leibnitz (FRM-II), TU M\"{u}nchen, D-85747 Garching, Germany}

\author{A.\ Hiess}
\affiliation{Institut Laue-Langevin, BP 156, F-38042 Grenoble, France}

\author{M.\ Boehm}
\affiliation{Institut Laue-Langevin, BP 156, F-38042 Grenoble, France}

\author{R.\ Mottl}
\affiliation{Laboratory for Neutron Scattering, ETH Zurich and PSI Villigen, CH-5232 Villigen PSI, Switzerland}

\author{S.\ Pailh\'es}
\affiliation{Laboratory for Neutron Scattering, ETH Zurich and PSI Villigen, CH-5232 Villigen PSI, Switzerland}

\author{N.\ Momono}
\affiliation{Department of Physics, Hokkaido University - Sapporo 060-0810, Japan}

\author{M.\ Oda}
\affiliation{Department of Physics, Hokkaido University - Sapporo 060-0810, Japan}

\author{M.\ Ido}
\affiliation{Department of Physics, Hokkaido University - Sapporo 060-0810, Japan}

\author{J.\ Mesot}
\affiliation{Laboratory for Neutron Scattering, ETH Zurich and PSI Villigen, CH-5232 Villigen PSI, Switzerland}
\affiliation{Laboratory for Quantum Magnetism, \'Ecole Polytechnique F\'ed\'erale de Lausanne (EPFL), CH-1015 Lausanne, Switzerland}

\begin{abstract}
Inelastic neutron-scattering experiments on the high-temperature superconductor 
La$_{1.855}$Sr$_{0.145}$CuO$_{4}$ reveal a magnetic excitation gap $\Delta$ that decreases 
continuously upon application of a magnetic field perpendicular to the CuO$_2$ planes. 
The gap vanishes at the critical field required to induce long-range incommensurate 
antiferromagnetic order, providing compelling evidence for a field-induced soft-mode 
driven quantum phase transition.
\end{abstract}

\maketitle

Driven by the continued theoretical focus on strong electron correlations and magnetism as a route to unconventional superconductivity~\cite{monthoux07},
the last two decades have seen tremendous efforts invested to characterize the momentum and energy dependence of magnetic fluctuations 
in cuprate high-$T_c$ superconductors. 
One of the most remarkable and encouraging results emerging from these studies is that upon entering the superconducting state optimally 
and overdoped hole~\cite{masonPRL92,yamadaPRL95,lakenature99,bourgesPHYSICAb95,daiPRB01} and electron-doped~\cite{yamadaPRL03} cuprates develop an excitation
gap. This phenomenon manifests itself as a complete suppression of all magnetic fluctuations below a material-dependent energy scale, sometimes referred to as the 
{\it spin gap}, which scales with $T_c$~\cite{yamadaPRL03,daiPRB08}. The correlation between superconductivity and low-energy spin fluctuations is, however, much 
less clear in underdoped cuprates, where linear scaling between the gap energy and T$_c$ breaks down~\cite{bourgesPHYSICAb95,daiPRB08,changPRL07}.

Particularly revealing studies of the excitation gap have involved the application of a magnetic field perpendicular to the CuO$_2$ planes. 
In the electron doped compound Nd$_{1.85}$Ce$_{0.15}$CuO$_{4}$ it has been demonstrated that the excitation gap decreases linearly with increasing 
magnetic field and extrapolates to zero at $H_{c2}$ -- the upper critical field for superconductivity \cite{motoyamaPRL06}. 
The situation is more complex in hole-doped La$_{2-x}$Sr$_x$CuO$_4$ (LSCO) where application of a magnetic field 
(i)  for $x\gtrsim 0.15$ tends to induce spectral weight below the zero-field gap ~\cite{lakescience01,gilardiEURO4,tranquadaPRBO4} and 
(ii) for $x\lesssim 0.15$ enhances \cite{lakeNAT02,chang_prb2008} the characteristic zero-field incommensurate (IC) stripe/spin density wave (SDW) order
\cite{tranquada1995,noteSDW} or even induces such order where none was present in zero field~\cite{khaykovichPRB05,chang_prb2008}.  
The latter observations have found formal expression in a Ginzburg-Landau (GL) model for competing SDW and superconducting orders \cite{demlerPRL01} 
which predicts the existence of a line of quantum critical points (QCP's) in the $T=0$ doping-field phase diagram, separating superconducting states with 
and without coexisting magnetic order (SC+SDW and SC, respectively, in the notation of Refs. \cite{demlerPRL01}).
However, one of the key expectations for continuous quantum phase transitions in general \cite{sachdevbook,kivelsonRMP03} and of the GL model 
\cite{demlerPRL01} in particular -- the existence of a field-induced soft mode in the spin excitation spectrum -- has never been firmly established.

In this Letter, we present an inelastic neutron scattering study of the low-energy spin fluctuations in La$_{1.855}$Sr$_{0.145}$CuO$_{4}$,
which in the absence of a magnetic field is a homogeneous, magnetically disordered superconductor with an excitation 
gap. 
We show that the gap decreases as the field-induced transition to a 
magnetically ordered SC+SDW state is approached, and tends to zero at the point where SDW order sets in.
Our discovery shows that the spectral properties of underdoped and optimally doped LSCO are smoothly connected and 
is strongly suggestive that the $T=0$ doping-field phase diagram of LSCO hosts a line of soft-mode driven SDW QCP's 
terminating, for $H=0$ T, near $x \approx 1/8$.

\begin{figure*}
\begin{center}
\includegraphics[width=0.99\textwidth]{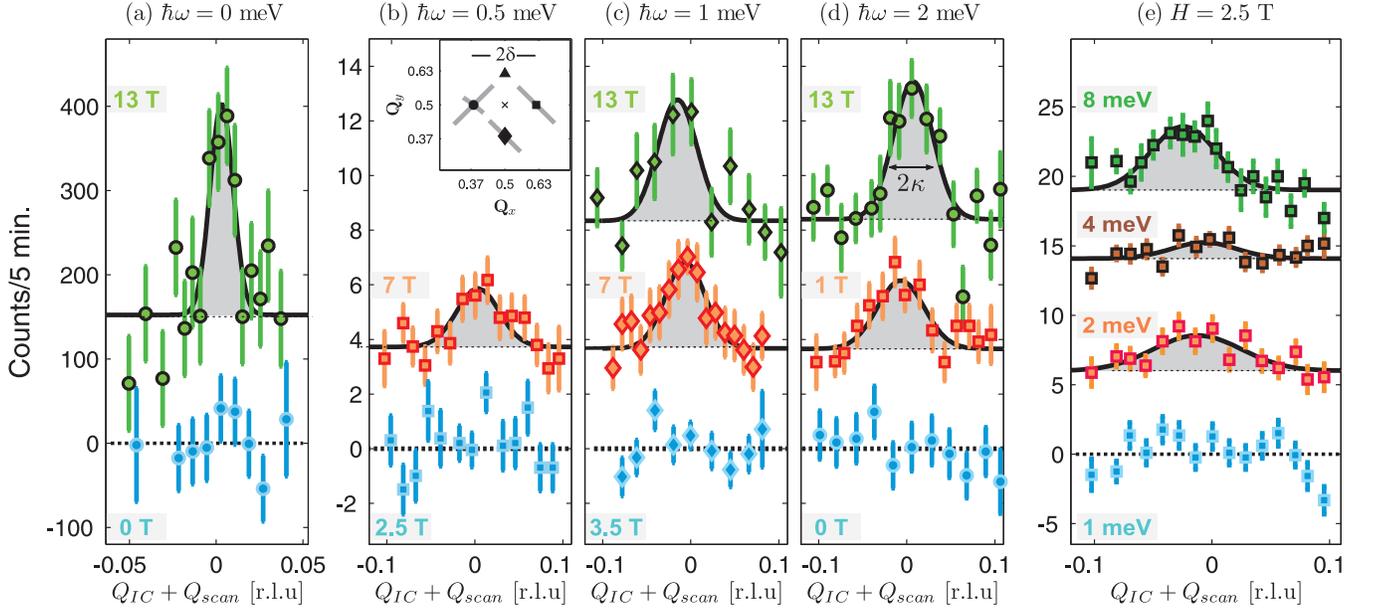}
\caption{
(a)-(e) Elastic and inelastic $\mathbf{Q}$-scans through $\mathbf{Q}_{IC}=(1/2\pm\delta,1/2,0), (1/2,1/2\pm\delta,0)$ as indicated in the inset of (b) \cite{note1}.
We use $\mathbf{Q}=\mathbf{Q}_{IC}+\mathbf{Q}_{\rm{scan}}$. 
(a) Elastic $\mathbf{Q}$-scans at external magnetic fields $0$~T and  $13$~T.
(b)-(d) Magnetic field dependence of $\mathbf{Q}$-scans recorded at fixed energy transfers $\hbar\omega=$ 0.5, 1 and 2 meV respectively.
(e) Energy dependence of $\mathbf{Q}$-scans obtained with $H=2.5$~T.
For clarity, the scans in (a)-(e) have been offset vertically with respect to each other.
The dashed lines are guides to the eye and the solid lines are Gaussian fits with a linear background. 
At $13$~T we find elastic and inelastic correlation lengths  $\xi($0 meV$)\sim150$~\AA\ and $\xi($2 meV$)\sim48$~\AA\ obtained from the half-width half-maximum 
$1/\kappa$ of the fitted peak.
}\label{fig:fig1}
\end{center}
\end{figure*}

The experiments were performed on the cold triple axis spectrometers 
IN14 at Institut Laue-Langevin, Grenoble, France and PANDA at FRM-II, 
Munich, Germany. On both instruments, cooled Be-filters were inserted after the sample to avoid higher-order contamination of the scattered beam of $5$~meV 
neutrons. 
The setup gave an energy resolution of 150~$\mu$eV (FWHM) or better, 
ensuring that there is no contribution from elastic scattering at the 
lowest energy transfers probed by our measurements. 
The sample consisted of two crystals (total mass $\approx 3.5$~g, $T_c\approx 36$~K), cut from the same travelling-solvent floating-zone grown \cite{TSFZ} 
rod, and co-aligned to within less than one degree. It was mounted in vertical field cryomagnets with the CuO$_2$ planes horizontal, allowing 
access to momentum transfers $(Q_h,Q_k,0)$. In tetragonal notation ($a \approx b = 3.81$ \AA, $c=13.2$ \AA), the low-energy 
magnetic response of superconducting LSCO peaks at a quartet of wavevectors $\mathbf{Q}_{IC}=(1/2\pm\delta,1/2,0), (1/2,1/2\pm\delta,0)$ as shown in the
inset of Fig.~\ref{fig:fig1}(b). 
We present data, recorded at $T \leq 3$~K, as a function of momentum $\mathbf{Q}$, energy transfer $\hbar\omega$ and external magnetic field $H$. 
Our results represent ground state properties of La$_{1.855}$Sr$_{0.145}$CuO$_{4}$ since the typical energies $\hbar\omega$ of the spin fluctuations 
studied are larger than the thermal energy $k_B T$. This also implies that measured peak amplitudes translate directly into magnetic susceptibility 
$\chi''(\mathbf{Q}_{IC},\omega)$ because the thermal population factor $[1-\exp(-\hbar\omega/k_BT)]^{-1}\approx 1$ is essentially irrelevant.

We start by noting that La$_{1.855}$Sr$_{0.145}$CuO$_{4}$ develops long-range magnetic order with $\delta\approx 0.13$ for $H > H_c = 7 \pm 1$~T~\cite{chang_prb2008},
as demonstrated by the appearance of sharp Bragg peaks at $\mathbf{Q}_{IC}$ which are absent for $H<H_c\ll H_{c2}$ (See Fig.~\ref{fig:fig1}(a)). 
Figures~\ref{fig:fig1}(b)-(d) show the variation with field of scans through $\mathbf{Q}_{IC}$ taken at 
$\hbar\omega=0.5$, $1$ and $2$ meV, respectively \cite{note1}. For all three energies,
the magnetic response is completely suppressed at the lowest fields shown (blue symbols),
but becomes finite and peaked at $\mathbf{Q}_{IC}$ upon application of higher fields 
(red and green symbols). 
Remarkably, in the case of $\hbar\omega=2$ meV, shown in Fig.~\ref{fig:fig1}(d), 
a field of just $1$ T is sufficient to induce an unambiguous magnetic excitation where none 
existed in zero field. Increasing the field leads to further enhancement of this signal.

Figures \ref{fig:fig1}(b)-(d) show that in moderate external fields
La$_{1.855}$Sr$_{0.145}$CuO$_{4}$ is non-responsive at low energy transfers, i.e. there is a gap 
in the magnetic excitation spectrum. In Fig.~\ref{fig:fig1}(e) we illustrate 
this point in a different manner, by plotting the $\hbar\omega$-dependence of 
scans through $\mathbf{Q}_{IC}$, all obtained at one fixed field, $H=2.5$ T. 
For $\hbar\omega = 1$ meV there is no magnetic response while the $\hbar\omega=$ 2, 4 and 8 meV scans
all display incommensurate peaks at $\mathbf{Q}_{IC}$. It is noteworthy that this magnetic signal 
has a non-monotonic energy-dependence with a minimum between 2 and 8 meV.

\begin{figure}
\begin{center}
\includegraphics[width=0.4\textwidth]{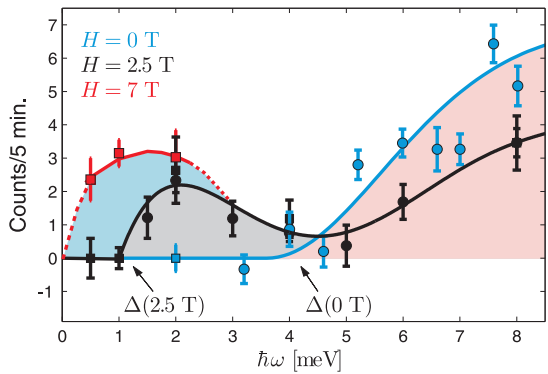}
\caption{Inelastic neutron scattering response at $\mathbf{Q}_{IC}$ as a function of energy transfer $\hbar\omega$
for $H=0$~T (blue), $H=2.5$~T (black) and $H=7$~T (red). The square points
are deduced from  Gaussian fits to $\mathbf{Q}$-scans as shown in Fig.~\ref{fig:fig1}
while circular points are from three point (background-$\mathbf{Q}_{IC}$-background)
scans. All lines are guides to the eye.}
\label{fig:fig2}
\end{center}
\end{figure}

Next, in Figure \ref{fig:fig2} we show the energy-dependence of the spectral weight at $\mathbf{Q}_{IC}$ for $H=0$, $2.5$ and $7$ T. 
In agreement with previous studies near optimal doping \cite{masonPRL92,yamadaPRL95,lakenature99}, the zero-field magnetic response is gapped. We define the
excitation gap $\Delta$ as the energy scale below which no spectral weight can be observed. The $H=0$ T measurement then yields 
$\Delta(0$~T$)=4\pm 0.5$~meV.
In a $2.5$~T field, complete suppression of spectral weight takes place only for $\hbar\omega<1.25$ meV, hence $\Delta(2.5$ T$)=1.25 \pm 0.5$~meV. 
As was clear already from Fig.~\ref{fig:fig1}(e), the  $2.5$ T spectrum displays noticeable local maxima and minima near $2$ meV and $4$ meV respectively. 
A comparison of the $0$ T and $2.5$~T spectra reveals that the field-induced spectral weight in the range $1.25 \lesssim \hbar\omega \lesssim 4$ meV has been 
transferred from energies above $\Delta(0$~T). 
Finally, at $7$~T, which is near the onset field for static SDW order, spin excitations are observed even at the lowest energy transfers probed
(See also Fig.~\ref{fig:fig1}(b)). This indicates that the gap has completely collapsed, $\Delta(7$ T$)= 0~\pm~0.5$~meV.

\begin{figure}
\begin{center}
\includegraphics[width=0.4\textwidth]{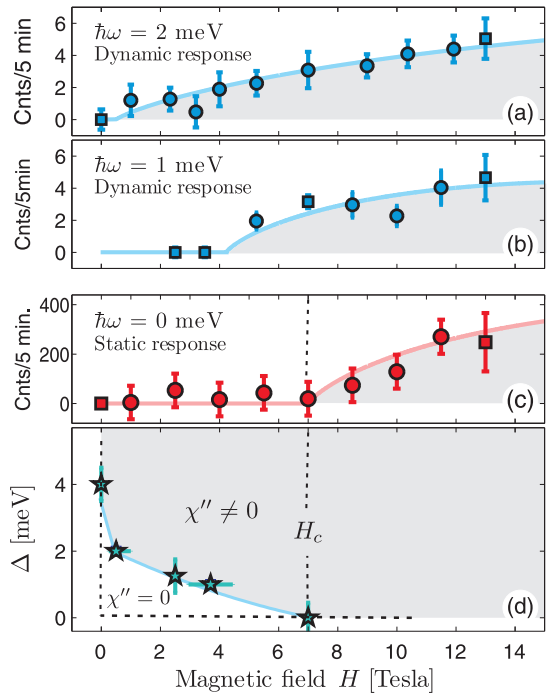}
\caption{
(a)-(b) Magnetic field dependence of the inelastic neutron response at $\mathbf{Q}_{IC}$ with
$\hbar\omega=2,$ and 1 meV, respectively. 
(c) Elastic neutron response at $\mathbf{Q}_{IC}$ as a function magnetic field. In (a)-(c) the square points
are deduced from  Gaussian fits to $\mathbf{Q}$-scans as shown in Fig.~\ref{fig:fig1}
while circular points are from three point (background-$\mathbf{Q}_{IC}$-background)
scans. All lines are guides to the eye. 
(d) Field-dependence of the excitation gap $\Delta$.}
\label{fig:fig3}
\end{center}
\end{figure}

The field-dependencies of the dynamic response at $\mathbf{Q}_{IC}$ for $\hbar\omega=2$ and $1$ meV are shown in 
Fig.~\ref{fig:fig3}(a) and (b), respectively.
For $\hbar\omega=2$~meV field-induced spin excitations are observed
already for $H=0.5 \pm 0.5$~T (See also Fig.~\ref{fig:fig1}(d)) and the magnetic 
signal increases continuously with increasing $H$.
By contrast, for $\hbar\omega=1$ meV the response is completely suppressed
for $H \lesssim 3.7 \pm 0.8$~T and excitations appear only at larger fields.
For comparison, Fig.~\ref{fig:fig3}(c) shows that the onset field for 
long-range static SDW order is $H_c =7 \pm 1$~T~\cite{chang_prb2008}.

We have now arrived at the main result of this Letter. Fig.~\ref{fig:fig3}(d) displays the field-dependence of the excitation gap $\Delta$, obtained 
by combining the data in Figs.~\ref{fig:fig2} and \ref{fig:fig3}(a)-(b). The gap is extremely sensitive to the application of a magnetic field: Following 
an initial dramatic drop, $\Delta$ subsequently softens more slowly and finally vanishes at the critical field $H_c\ll H_{c2}$ which marks 
the onset of long-range incommensurate magnetic order.

We start our discussion by pointing out a strong resemblance of the
field-induced spectral evolution in La$_{1.855}$Sr$_{0.145}$CuO$_{4}$, to that which takes place when the SC to SC+SDW
transition is approached through changes in chemical composition. 
Increasing levels of Zn replacing Cu in La$_{1.85}$Sr$_{0.15}$Cu$_{1-y}$Zn$_y$O$_4$, which for $y=0$ has a well-developed gap,
leads to suppression of $T_c$ and to a gradual shift of $\Delta$~\cite{kimura_prl2003}. 
Eventually, for $y=0.017$, long-range SDW order co-existing with SC sets in and no excitation gap can be resolved~\cite{kimura_prl2003}.
Similarly, as a function of decreasing Sr content, La$_{2-x}$Sr$_{x}$CuO$_{4}$ evolves from a SC state with $\Delta \sim 4$ meV, for $x>0.13$, 
to a SC+SDW state with significant peaked spectral weight below this energy scale~\cite{kofu}.
What we have discovered is that the same spectral evolution can be accomplished continuously by application of a magnetic 
field and with no variation in chemical disorder.

One of the hallmarks of a continuous quantum phase transition is the 
existence of a mode in the excitation spectrum which responds to changes in an experimentally 
tunable parameter (such as magnetic field, pressure or doping) by softening towards zero, causing qualitative 
changes in the ground state wavefunction once the $\hbar\omega=0$ limit is reached ~\cite{sachdevbook,kivelsonRMP03}.
Quantum disordered spin dimer systems have proven to be a fertile ground for studies of this canonical behaviour.
For example, in the case of TlCuCl$_3$, the Zeeman splitting of excited triplet states in a magnetic field 
leads to a linear reduction of the singlet-triplet gap and to Bose-Einstein condensation
of magnons and gapless spin excitations above the critical field required to fully close the gap~\cite{rueggNAT03}.
By analogy, we interpret our observation of a gradual (although clearly non-linear at small $H$) reduction 
of $\Delta$ and closure of this gap at the onset field for magnetic order, as evidence that the 
field-induced SC to SC+SDW transition is a continuous quantum phase transition. An early demonstration
of nearly singular magnetic fluctuations and scaling \cite{aeppliscience97} in a sample of almost identical 
composition to ours provides further support for this interpretation.

The observed field-dependence of $\Delta$ is in qualitative agreement with a key 
prediction of GL theory for SC to SC+SDW quantum phase transitions \cite{demlerPRL01}, 
namely the appearance of a field-induced soft precursor mode 
(a $S=1$ collective "spin resonance" \cite{demlerPRL01}, associated with 
oscillations of the SDW order parameter about zero) centered, 
for $H\! \leq \!H_{c}\! \leq \! H_{c2}$, at energy $\epsilon(H)=\epsilon(0)+C_1(H/H_{c2})\log (H_{c2}/H)$ 
with $\epsilon(0)$ and $C_1$ being constants. To make the connection to GL theory, we must interpret $\Delta$ as marking 
the low-energy tail of the soft mode. In turn, this suggests to view the low-energy peak in the complex spectral lineshape at 
$2.5$ T (See Fig.~\ref{fig:fig2}) as the soft mode.
The existence of a field-induced spectral peak was first reported in LSCO $x=0.16$ \cite{lakescience01}. 
Subsequent experiments on $x=0.17$~\cite{gilardiEURO4} and $x=0.18$~\cite{tranquadaPRBO4} samples gave 
no indications of a well-defined mode, although in-gap spectral weight was reported in both cases. Our results 
reafirm the existence of a field-induced mode in the SC phase of LSCO and suggest, as would be implied by the conditions 
of validity of the GL model \cite{demlerPRL01}, that it is most easily observed when in close proximity to the line of 
continuous SC to SC+SDW quantum phase transitions.

More generally, closure of the excitation gap appears to be a universal phenomenon in cuprates 
hosting quantum phases in which superconductivity can coexist with magnetic order. In YBa$_2$Cu$_3$O$_{6+p}$, 
the gap collapses abruptly for $p\approx0.5$ \cite{daiPRB08}, and a robust SC phase 
with incommensurate, quasi-static fluctuations (an electronic nematic) 
emerges at lower doping levels \cite{hinkov}. 
On the other hand, in Nd$_{1.85}$Ce$_{0.15}$CuO$_{4}$ where SC and magnetism do not coexist, the excitation gap extrapolates to zero only at 
$H_{c2}$ \cite{motoyamaPRL06}.

These considerations raise the possibility that when coexistence of superconductivity and magnetism is an issue,
two separate energy scales need to be considered: (i) The zero-field gap $\Delta(0$ T$)$ -- related to 
superconductivity and (ii) a gap related to magnetic order. Further, these scales may display distinct 
field-dependencies. 
Related ideas about spectral separation were presented in Ref. \cite{kofu} in the context of spatial phase 
separation between SC and magnetically ordered regions.
By not requiring $\epsilon(0)=\Delta(0$ T$)$, GL theory has spectral separation built in with
no need for phase separation beyond what is implied by the soft mode being stabilized near vortices 
\cite{demlerPRL01}. 
Interestingly, Quantum Monte Carlo  computations of the excitation spectrum of a mixture of magnetically 
ordered and disordered patches, intended to model the effect of SDW order pinned by vortices or impurities, 
does produce a low-energy peak below the excitation gap of the fully disordered system \cite{bma}. 
Further experimental and theoretical work is needed to clarify these issue.

In summary, we have discovered that under the influence of a magnetic field applied perpendicular to the 
CuO$_2$ planes, the gap to magnetic excitations in La$_{1.855}$Sr$_{0.145}$CuO$_4$ decreases gradually 
and vanishes at the onset field for long-range static SDW order. 
We have argued that these observations, which follow the expectations for a continuous soft-mode driven 
quantum phase transition, suggest the existence of a line of such transitions in the doping-field phase
diagram of LSCO.

This work was supported by the Swiss NSF (through NCCR, MaNEP, and grant Nr 200020-105151, PBEZP2-122855) and by the Ministry
of Education and Science of Japan. We gratefully acknowledge discussions with C. Mudry, B. M. Andersen, M. Kenzelmann, A.-M. Tremblay, B. Keimer, 
L. Taillefer and A. T. Boothroyd.


\begin{thebibliography}{99}

\bibitem{monthoux07}
P.Monthoux \textit{et al.}, Nature \textbf{450}, 1177 (2007).

\bibitem{masonPRL92}
T. E. Mason \textit{et al.}
Phys. Rev. Lett. \textbf{68}, 1414 (1992).

\bibitem{yamadaPRL95}
K. Yamada \textit{et al.}
Phys. Rev. Lett. \textbf{75}, 1626 (1995).

\bibitem{lakenature99}
B. Lake \textit{et al.}
Nature \textbf{400}, 43 (1999).

\bibitem{bourgesPHYSICAb95}
P. Bourges \textit{et al.}
Physica B (Amsterdam) \textbf{215}, 30 (1995).

\bibitem{daiPRB01}
P. Dai \textit{et al.}
Phys. Rev. B \textbf{63}, 054525 (2001).

\bibitem{yamadaPRL03}
K. Yamada \textit{et al.}
Phys. Rev. Lett. \textbf{90}, 137004 (2003).

\bibitem{daiPRB08}
S. Li \textit{et al.}
Phys. Rev. B \textbf{77}, 014523 (2008).

\bibitem{changPRL07}
J. Chang \textit{et al.}
Phys.~Rev.~Lett.~\textbf{98}, 077004 (2007).

\bibitem{motoyamaPRL06}
E. M. Motoyama \textit{et al.}
Phys. Rev. Lett. \textbf{96}, 137002 (2006).

\bibitem{lakescience01}
B. Lake \textit{et al.}
Science \textbf{291}, 1759 (2001).

\bibitem{gilardiEURO4}
R. Gilardi \textit{et al.}
Europhysics Letters \textbf{66}, 840 (2004).

\bibitem{tranquadaPRBO4}
J. Tranquada \textit{et al.}
Phys. Rev. B \textbf{69}, 174507 (2004).

\bibitem{chang_prb2008}
J. Chang, \textit{et al.}
Phys. Rev. B \textbf{78}, 104525  (2008).

\bibitem{lakeNAT02}
B. Lake \textit{et al.}
Nature \textbf{415}, 299 (2002).

\bibitem{tranquada1995}
J. M. Tranquada \textit{et al.}
Nature \textbf{375}, 561 (1995).

\bibitem{noteSDW}
In this paper, we make no distinction between stripe~\cite{tranquada1995} and spin density wave order.
The abbreviation SDW is used for any magnetic order having dominant 
Fourier components at the incommensurate wavevectors $\mathbf Q_{IC}$.

\bibitem{khaykovichPRB05}
B. Khaykovich \textit{et al.}
Phys. Rev. B \textbf{71}, 220508(R) (2005).

\bibitem{demlerPRL01}
E. Demler \textit{et al.} Phys. Rev. Lett. \textbf{87}, 067202 (2001);
Y. Zhang  \textit{et al.} Phys. Rev. B \textbf{66}, 094501 (2002).

\bibitem{sachdevbook}
S. Sachdev, {\it Quantum Phase Transitions}, (Cambridge. University Press, Cambridge, 1999).

\bibitem{kivelsonRMP03}
S. A. Kivelson \textit{et al.}
Rev. Mod. Phys. \textbf{75}, 1201 (2003).

\bibitem{TSFZ}
T. Nakano \textit{et al.}
J. Phys. Soc. Jpn. \textbf{67}, 2622 (1998).

\bibitem{note1}
We have no reasons to suspect that the dynamic response at the four peaks 
should be different in the magnetically disordered state since it must 
respect the pseudo-tetragonal crystal structure~\cite{kivelsonRMP03}. 
In each experiment we therefore chose to study the peak that 
had the best signal-to-noise ratio at $2$~meV in zero field at $40$ K. 

\bibitem{kimura_prl2003}
H. Kimura \textit{et al.}
Phys. Rev. Lett. \textbf{91}, 067002 (2003). 

\bibitem{kofu}
M. Kofu \textit{et al.}
arXiv:0804.3041

\bibitem{rueggNAT03}
C. R\"uegg \textit{et al.}
Nature \textbf{423}, 62 (2003). 

\bibitem{aeppliscience97}
G. Aeppli \textit{et al.}
Science \textbf{278}, 1432 (1997).

\bibitem{hinkov}
V. Hinkov \textit{et al.}
Science \textbf{319}, 597 (2008).

\bibitem{bma}
B. M. Andersen, O. Sylju{\aa}sen and P. Hedeg{\aa}rd, to be published.

\end{thebibliography}
\end{document}